\documentstyle[12pt,aaspp4,epsfig,flushrt,natbib]{article}
\begin{document}

\title{Discovery of a 450 Hz QPO from the Microquasar GRO J1655-40 with RXTE}
\author{Tod E. Strohmayer}
\affil{Laboratory for High Energy Astrophysics, NASA's Goddard Space Flight 
Center, Greenbelt, MD 20771; stroh@clarence.gsfc.nasa.gov}

\begin{abstract}

We report the discovery with the proportional counter array (PCA) onboard the 
Rossi X-ray Timing Explorer (RXTE) of a 450 Hz quasiperiodic oscillation (QPO)
in the hard X-ray flux from the galactic microquasar GRO J1655-40. This is the
highest frequency QPO modulation seen to date from a black hole. The QPO 
is detected only in the hard X-ray band above $\sim 13$ keV. 
It is both strong and narrow, with a typical rms amplitude of 4.5 \% in the 
13 - 27 keV range, and a width of $\sim 40$ Hz (FWHM). For two observations in 
which we detect the 450 Hz QPO a previously known $\sim 300$ Hz QPO is 
also observed in the 2 - 13 keV band. We show that these two QPO sometimes 
appear simultaneously, thus demonstrating the first detection of a 
pair of high frequency QPO in a black hole system. Prior to this, pairs of high
frequency QPO have only been detected in neutron star systems. GRO J1655-40 is 
one of only a handful of black hole systems with a good dynamical mass 
constraint. For a non-rotating black hole with mass between 5.5 - 7.9 
$M_{\odot}$ the innermost stable circular orbit (ISCO) ranges
from 45 - 70 km. For any mass in this range the radius at which the orbital 
frequency reaches 450 Hz is less than the ISCO radius, indicating that if the 
modulation is caused by Kepler motion, the black hole must have
appreciable spin. If the QPO frequency is set by the orbital frequency of 
matter at the ISCO then for this mass range the dimensionless angular momentum 
lies between $0.15 < j < 0.5$. Moreover, if the modulation is caused by 
oscillation modes in the disk or Lense-Thirring precession, then this would 
also require a rapidly rotating hole. We briefly discuss the implications of 
our findings for models of X-ray variability in black holes and neutron stars.

\end{abstract}


\centerline{Accepted for publication in Astrophysical Journal Letters}

\vfill\eject

\section{Introduction}

The Galactic microquasar GRO J1655-40 is one of a handful of black hole systems
in which high frequency quasiperiodic X-ray brightness oscillations have been 
recently observed with RXTE. Remillard et al. (1999) reported the presence of a
weak, broad $\sim 300$ Hz QPO at times when the X-ray spectrum was hardest 
(dominated by a power law component) and the luminosity above $\sim 0.2 
L_{Edd}$. Recently, high frequency QPOs have also been observed in three black 
hole transients; 4U 1630-47 at 184 Hz, XTE J1858+226 at 170 Hz, and XTE 
J1550-564 between 100 - 283 Hz (see Remillard \& Morgan 1998; Markwardt et al. 
1999; Remillard et al. 1999a; Homan et al 2000). A stable 67 Hz QPO has also 
been observed from the Galactic microquasar GRS 1915+105 (Morgan, Remillard
\& Greiner 1997). The high frequencies 
of these QPO combined with their X-ray origin argue forcefully that 
they are produced in the innermost region of the accretion flow close to the 
black hole event horizon. If their production mechanism can be understood
they stand to provide a wealth of information on black hole mass and spin as 
well as the structure of strongly curved spacetime. Their behavior may also
provide tests of General Relativity (GR) in the strong field limit. A number
of different models have been proposed to explain these oscillations, but all 
of them require that strong-field GR effects be taken into 
account (see Milsom \& Taam 1997; Nowak et al. 1997; Wagoner 1998; Stella, 
Vietri \& Morsink 1999; Merloni et al. 1999). Although pairs of simultaneously
present high frequency QPOs have been observed ubiquitously from neutron star
systems (see the review by van der Klis 2000), until now no black hole system 
has revealed a pair of simultaneous high frequency ($\nu \ge 50$ Hz) QPO. 

GRO J1655-40 is unique amongst these QPO sources in that it is the only one 
whose mass is known with reasonable precision.  
Orosz \& Bailyn (1997) observed the source in quiescence, modelled the 
observed ellipsoidal variations using both quiescent and outburst data 
and obtained a mass $M = 7.02 \pm 0.22 
M_{\odot}$. More recently, Shabaz et al. (1999) used radial velocity
observations in quiescence to constrain the black hole mass to the range 
$5.5 - 7.9 M_{\odot}$. Since these results were obtained using only quiescent
data they should be free from systematic effects due to X-ray 
heating of the secondary which can bias the mass determination (Phillips, 
Shahbaz \& Podsiadlowski 1999). These mass constraints make GRO J1655-40 
arguably the best object in which to test different models for the high 
frequency X-ray oscillations and thus to learn how to infer fundamental 
properties of black holes from X-ray variability measurements. 

In this Letter we report the discovery in archival RXTE data of a strong and
narrow $\sim 450$ Hz QPO in the hard X-ray flux from GRO J1655-40. We show that
this QPO is present at the same time as a previously discovered 300 Hz QPO
(Remillard et al. 1999), however, the 450 Hz QPO is not always seen when the 
300 Hz QPO is detected and vice versa. This result is the first detection
of a {\it pair} of high frequency QPO in a black hole system, and is also the 
highest frequency modulation yet seen from a black hole. We discuss the
implications of our findings for models of X-ray variability in neutron star 
and black holes systems as well as the implications for black hole spin in 
GRO J1655-40.

\section{Data Analysis}

GRO J1655-40 is one of the best studied Galactic `microquasars', a class of
black hole binaries which produce superluminal radio jets and correlated 
X-ray activity (Mirabel \& Rodriquez; Hjellming \& Rupen 1994). The source
was discovered by BATSE in July, 1994 (Harmon et al. 1995). 
A new outburst of GRO J1655-40 was discovered in April, 1996 by the All-Sky
Monitor (ASM) aboard RXTE (Remillard et al. 1996). The outburst lasted more
than 450 days and an extensive set of RXTE pointed observations were conducted.
Remillard et al. (1999) reported an extensive timing analysis of 52 RXTE 
observations. They reported four types of QPO behavior spanning the range 
from 0.1 - 300 Hz. They found that the 300 Hz QPO as well as a 0.1 Hz QPO 
appeared only when the X-ray spectrum was dominated by a hard power law 
component. 

We began investigating the X-ray timing properties of GRO J1655-40 as part of 
an effort to compare systematically the variability properties of black holes 
and neutron stars in the RXTE archive. The data
we discuss here were obtained from August through November, 1996 as part of 
proposal P10255 (PI: Remillard) and are now public. The data modes employed for
most of these observations included a high time resolution (sampling rate of 
65,536 Hz) event mode recording events above PCA channel 35 ($\sim 13$ keV) 
and a single bit mode covering the lower energies in a single band from 2 - 12 
keV. As part of our analysis we computed power spectra for each observation
and data mode separately. We used 32 s intervals and 8 Hz frequency 
resolution in the range from 1 - 2048 Hz. We then averaged the individual 
spectra for each observation. We found almost immediately that some of the 
average power spectra computed using only the high energy event mode data
contained a significant feature near 450 Hz. Figure 1 shows the 
power spectrum for observation 10255-01-10-00 
(UT date, 09/09/96, Observation 13 in Table 2 of Remillard et al. 1999) in 
which we first saw the 450 Hz feature.

To estimate the significance of this feature we first rescaled 
the power spectrum so that the local mean near 450 Hz was 2 (the value 
expected for a Poisson process), and then computed the probability of 
obtaining a power $P = P_{max}\times 512\times 256$ from a $\chi^2$ 
distribution with $2\times 512\times 256$ degrees of freedom. 
Here $P_{max}$ is the highest power in the QPO feature. 
We used this $\chi^2$ distribution because we averaged in 
frequency by a factor of 512 and averaged 256 individual power spectra. 
This gives a chance probability of $2\times 10^{-8}$ for the highest bin 
within the QPO profile, better than a 5$\sigma$ deviation. We note that this 
is also a conservative estimate since the QPO feature has not been averaged 
into a single bin at this frequency resolution. 

We next modelled the power spectrum using a Gaussian profile and a power law
component in the 100 - 1200 Hz range. The QPO is well fit by a Gaussian 
centered at $\nu_0 = 449.3 \pm 5$ Hz and a width of $\delta\nu = 19.6 
\pm 4$ Hz. This gives a coherence value $Q = \nu / \delta\nu = 23$. The rms
amplitude in the 13 - 27 keV band is $4.8 \pm 0.6 \%$. 

Having found a significant QPO in one of the observations we then searched the 
remaining set of observations for a similar feature. We detected the 450 Hz QPO
in four additional observations (10255-01-06-01 on 8/16/96, 10255-01-07-00 on
8/22/96, 10255-01-09-00 on 9/4/96, and 10255-01-17-00 on 10/27/96). Power 
spectra from the three observations with the strongest detections are shown in 
figure 2 (left panel). We modelled the QPO in each observation in which it was 
detected and found that its centroid did not change significantly over those 
observations. Thus, to within the precision of the measurements the frequency 
appears to be stable. Finally, we computed an average power spectrum for all 
observations which contained the 450 Hz QPO (see figure 2). 

\section{Discussion and Summary}

The detection of a 450 Hz QPO in GRO J1655-40 has many
important implications and also raises many interesting questions. One of the 
most important questions bears on the relationship, if any, between the
300 Hz QPO discovered by Remillard et al. (1999) and the 450 Hz QPO reported
here? We detected the 450 Hz QPO in three observations in which Remillard et 
al. (1999) detected the 300 Hz QPO; 10255-01-06-01 on 8/16/96, 10255-01-07-00 
on 8/22/16 and 10255-01-17-00 on 10/27/96. Note that for the latter 
observation, Remillard et al. report a possible occurrence of the 300 Hz QPO. 
To investigate whether both QPOs can appear simultaneously
we computed an average power spectrum by combining these three observations. We
computed two power spectra, one using only the 2 - 12 keV single bit data, and 
the other using just the 13 - 27 keV event data. Our resulting power spectra 
are shown in figure 3. We clearly detect {\it both} the 300 Hz QPO in the 
2 - 12 keV band and the 450 Hz QPO in the hard band. We also investigated the
average power spectra of each observation individually. Both the 300 Hz and 
450 Hz QPOs are detected in the 8/16/96 and 8/22/96 data. The 450 Hz QPO is 
detected in the 10/27/96, but the 300 Hz QPO is only marginally 
detected, consistent with the findings of Remillard et al. (1999). During the 
8/22/96 observation we were able to detect the 450 Hz QPO in several 
subintervals and we confirm that during this observation the 450 Hz frequency
did not drift significantly. Although in this observation we can not track the 
frequency of the 300 Hz QPO with the same temporal resolution as the 450 Hz 
QPO, it's stability on timescales of several hours is evident in other 
observations (for example 10255-01-04-00 on 8/1/96). This evidence, combined 
with the observed sharpness of both QPOs in the average power spectrum (see
figure 3), provides strong evidence that the pair of peaks could not be 
produced by a single feature which drifted substantially in frequency. 

Remillard et al. (1999) report 300 Hz QPO in three additional observations in 
which we did not detect the 450 Hz QPO. We also investigated the average power
spectra of these observations. The results confirm the 300 Hz detections but 
also show no evidence for the 450 Hz QPO. This indicates that the two QPOs are
not {\it always} detected together. Interestingly, the pairs of kilohertz QPO 
seen in neutron star systems are also not always detected at the same time 
(see van der Klis 2000). 

The 300 Hz QPO has a typical amplitude in the 2 - 12 keV band of $\sim
0.8 \%$. Such an amplitude would not be detectable in the 13 - 27 keV
data. If the amplitude increased to about $1.7 \%$ in the hard band then we
would have been able to just detect the signal, so we can place a limit on
the amplitude of the 300 Hz signal in the 13 - 27 keV band of about $1.7 \%$.
Similarly, if the 450 Hz QPO amplitude stayed the same down into the soft
band then we would have detected it easily, so its amplitude must be a strong
function of energy, and in fact, it cannot be much larger than about $0.4\ \%$
in the 2 - 12 keV band or it would have been detected. Not only are two 
high frequency signals present but they appear to have very different energy 
dependencies as well, perhaps indicating different physical mechanisms or
production sites for each oscillation. 

The detection of a 450 Hz QPO in GRO J1655-40 has interesting implications for
black hole spin in this source. A black hole whose mass lies between 
$5.5 - 7.9 M_{\odot}$, the 95 \% confidence limits from Shahbaz et al. (1999), 
must have a {\it non-zero} angular momentum in order for the orbital frequency 
at the ISCO to be greater than or equal to 450 Hz (see Bardeen, Press \& 
Teukolsky 1972 for a discussion of the ISCO radius). In figure 4a (left) we 
have plotted the ISCO radii as a function of dimensionless angular momentum 
$j = cJ /GM^2$ using the mass limits for GRO J1655-40 from Shahbaz et al. 
(1999) and Orosz \& Bailyn (1997). For each mass limit we also show the 
corresponding radii at which the orbital frequency would be 450 Hz. Even at the
lower mass limit of Shahbaz et al. (1999) one requires $j \ge 0.15$ in order 
for the orbital frequency at the ISCO to be 450 Hz. Indeed, if the 450 Hz QPO 
is set by the orbital frequency at the ISCO, then $j$ could range from about 
0.15 to 0.5. Only with $M \le 4.8 M_{\odot}$ would a non-rotating hole be 
able to produce a 450 Hz orbital frequency at the ISCO.

Although the 450 Hz QPO may not actually be produced by Kepler motion of 
matter at some radius, the Kepler frequency is generally the highest 
characteristic variability frequency at a given radius. For example, the 
lowest order diskoseismic modes 
discussed by Perez, Silbergleit \& Wagoner (1999), some of which have been
suggested as the mechanism for high frequency black hole QPOs (Nowak et al.
1997), all have characteristic frequencies {\it below} that of the Kepler 
frequency at the ISCO. Therefore, if one of these modes is responsible for 
the 450 Hz oscillation the conclusion that the black hole has appreciable spin 
is unchanged, in fact, in such a case it would have to be spinning even 
faster than suggested by the orbital frequency at the ISCO. Moreover, 
Cui, Zhang, \& Chen (1998) suggested that the 300 Hz QPO in GRO J1655-40 could
be associated with the Lense-Thirring precession frequency close to the ISCO 
radius. For GRO J1655-40 this could only be achieved with a nearly maximally 
rotating black hole. With the discovery of a 450 Hz QPO it is now even less 
obvious which frequency should be associated with Lense-Thirring precession 
(see Mendez, Belloni \& van der Klis 1998). If 300 Hz is indeed the precession 
frequency, then the 450 Hz QPO cannot be due to Kepler motion at the same 
radius as this would have a much higher frequency. However, if either of the 
QPO can be associated with Lense-Thirring precession the black hole {\it must} 
be nearly maximally rotating. Although controversial, spectral analysis by 
Zhang, Cui \& Chen (1997) suggests that the inner accretion disk of 
GRO J1655-40 is hot and compact, perhaps requiring rapid black hole spin 
(see however, Merloni, Fabian \& Ross 2000; Sobczak et al. 1999). 

There has been much recent work directed at investigating the relationship 
between high frequency QPOs and characteristic noise frequencies
in neutron stars and black holes. For example, Psaltis et al. (1999) showed 
that a correlation could be found between a pair of QPO frequencies across a 
broad range of source types and luminosities, including both neutron star and 
black hole systems. In the neutron star systems the pair of frequencies are the
lower frequency kilohertz QPO and the 20 - 50 Hz QPO most commonly referred 
to as a horizontal branch oscillation (HBO). For the black hole systems the 
situation is a bit less certain because of the lack of high frequency features 
with properties unambiguously similar to the neutron star kHz QPO. In the 
so called relativistic precession models the observed QPO frequencies have been
identified with the Keplerian, the periastron precession and nodal precession
frequencies at some characteristic radius in the accretion disk (see Stella, 
Vietri \& Morsink 1999; Psaltis \& Norman 2000; Markovic 2000; Markovic \& Lamb
2000). Psaltis et al. (1999) tentatively identified the 300 Hz QPO from 
GRO J1655-40 with the lower kHz QPO in neutron stars. Our detection of a 
higher frequency QPO in GRO J1655-40 (perhaps the analog to the upper kHz QPO
in neutron stars?) would at first glance seem to strengthen this association. 
To test this we investigated whether the pair of high frequency QPO and the 
lower frequency ($\sim 18$ Hz) QPO could be consistently
associated with the Keplerian, periastron precession and nodal precession 
frequencies of a $\sim 7 M_{\odot}$ black hole. We show in figure 4b a plot of 
the radial epicyclic frequency vs the Kepler frequency for the mass range  
appropriate for GRO J1655-40 and several different values of the dimensionless 
angular momentum $j$ (curved traces). In the relativistic models the frequency 
{\it difference} between the pair of high frequency QPOs is identified with the
radial epicyclic frequency. We also plot the nodal precession frequency for 
the same set of angular momenta and mass range. The QPO data from GRO J1655-40 
are shown with diamond symbols. A black hole with $0.4 < j < 0.6$ could account
for the high frequency QPOs, but then the predicted nodal precession frequency 
is uncomfortably high to be associated with the $\sim 18$ Hz QPO. Whether or 
not hydrodynamic corrections can mitigate this apparent discrepancy remains to 
be seen (see Psaltis 2000).

Models for the Khz QPO in neutron star systems generally fall into two broad
classes; beat frequency models (Miller, Lamb, \& Psaltis 1998; Strohmayer
et al. 1996), which require the spinning neutron star surface to produce
a pair of frequencies; and relativistic disk models (Stella, Vietri, \& Morsink
1999; Psaltis \& Norman 2000; Nowak et al. 1997; Perez et al. 2000), which
generate the frequencies in the disk alone and are therefore thought to be
more generally applicable to both neutron star and black hole systems. Our
discovery of a 2nd high frequency QPO in GRO J1655-40 proves that it is not
necessary to have a hard surface (as in a neutron star) to produce a pair
of high frequency peaks. This would seem to provide some support to the
relativistic disk models, since this model can naturally produce pairs of 
peaks in either class of source, however, a straightforward application of the
model seems to have some difficulty (see above). Caution regarding the
identification of the QPOs in GRO J1655-40 with the better understood
kHz QPO in neutron stars is warranted because of the different properties 
of the QPOs in GRO J1655-40 and the neutron star kHz QPOs.


\acknowledgements

We thank Craig Markwardt, Cole Miller and Jean Swank for many helpful 
discussions and comments on the manuscript. We thank the referee, Ron 
Remillard, for his comments which helped us improve the paper. 

\vfill\eject

\section*{References}

\noindent Bardeen, J. M., Press, W. H. \& Teukolsky, S. A. 1972, ApJ, 178, 347

\noindent Cui, W., Zhang, S. N. \& Chen, W. 1998, ApJ, 492, L53

\noindent Harmon, B. A. et al. 1995, Nature, 374, 703

\noindent Hjellming, R. M. \& Rupen, M. P. 1995, Nature, 375, 464

\noindent Homan, J. et al. 2000, ApJ in press, (astro-ph/0001163)

\noindent Leahy, D. A. et al. 1983, ApJ, 266, 160

\noindent Markovic, D. 2000, MNRAS, submitted, (astro-ph/0009450)

\noindent Markovic, D. \& Lamb, F. K. 2000, MNRAS, submitted, 
(astro-ph/0009169)

\noindent Markwardt, C.\ B., Swank, J.\ H.\ \& Taam, R.\ E.\ 1999, ApJ, 513, 
L37

\noindent Mendez, M., Belloni, T., \& van der Klis, M. 1998, ApJ, 499, L187

\noindent Merloni, A., Fabian, A. C. \& Ross, R. R. 2000, MNRAS, 313, 193

\noindent Merloni, A., Vietri, M., Stella, L. \& Bini, D. 1999, MNRAS, 304, 155

\noindent Miller, M. C., Lamb, F. K. \& 
Psaltis, D. 1998, ApJ, 508, 791

\noindent Milsom, J. A., \& Taam, R. E. 1997, MNRAS, 286, 359

\noindent Mirabel, I. F. \& Rodriguez, L. F. 1994, Nature, 371, 46

\noindent Morgan, E. H., Remillard, R. A. \& Greiner, J. 1997, ApJ, 482, 993

\noindent Nowak, M. A., Wagoner, R. V. Begelman, 
M. C. \& Lehr, D. E. 1997, ApJ, 477, L91

\noindent Orosz, J. A. \& Bailyn, C. D. 1997, ApJ, 477, 876

\noindent Perez, C. A., Silbergleit, A. S., Wagoner, R. V. \& Lehr, D. E. 
1997, ApJ, 476, 589

\noindent Phillips, S. N., Shahbaz, T. \& Podsiadlowski, P. 1999, MNRAS, 304, 
839

\noindent Psaltis, D. 2000 ApJ, submitted, (astro-ph/0010316)

\noindent Psaltis, D., Belloni, T. \& van der Klis, M. 1999, ApJ, 520, 262

\noindent Psaltis, D. \& Norman, C. 2000, ApJ, in press (astro-ph/0001391)

\noindent Remillard, R. A., et al. 1996, IAU Circ. 6393

\noindent Remillard, R. A., Morgan, E. H., McClintock, J. E., Bailyn, C. D. 
\& Orosz, J. A. 1999, ApJ, 522, 397

\noindent Remillard, R. A. \& Morgan, E. H. 1998, in The Active X-ray Sky, ed. 
L. Scarsi, H. Bradt, P. Giommi, \& F. Fiore (Amsterdam: Elsevier), 316

\noindent Remillard, R. A., Morgan, E. H., 
McClintock, J. E., Bailyn, C. D. \& Orisz, J. A. 1998, in Proc. 18th Texas
Symp. on Relativistic Astrophysics, ed. A. Olinto, J. Frieman, \& D Schramm 
(Singapore: World Scientific), 750

\noindent Shahbaz, T., van der Hooft, F., Casares, J., Charles, P. A. \& van 
Paradijs, J. 1999, MNRAS, 306, 89

\noindent Sobczak, G. J., McClintock, J. E., Remillard, R. A., Bailyn, C. D. 
\& Orosz, J. A. 1999, ApJ, 520, 776

\noindent Stella, L., Vietri, M. \& Morsink, S. M. 1999, ApJ, 524, L63

\noindent Strohmayer, T. E., Zhang, W., Swank, J. H., Smale, A. P., Titarchuk, 
L., Day, C. \& Lee, U. 1996, ApJ, 469, L9

\noindent van der Klis, M. 2000, ARAA, in press (astro-ph/0001167)

\noindent Zhang, S. N., Cui, W. \& Chen, W. 1997, ApJ, 482, L155


\vfill\eject

\section{Figure Captions}

Figure 1: Average power spectrum for observation 
10255-01-10-00 in the 13 - 27 keV energy band. The spectrum was computed by
averaging a total of 256 individual spectra each computed from a 32 second
interval. The frequency resolution is 16 Hz and the spectrum has been 
normalized following Leahy et al. (1983). The QPO at 450 Hz is clearly
visible.

\vskip 10pt

Figure 2: Average power spectra from three
observations with the most significant detections of the 450 Hz QPO (top 
three traces). The spectra were computed in the same way as for figure 1. The 
centroid of the 450 Hz peak did not change significantly over these 
observations. The spectra have been shifted vertically for clarity.
The average power spectrum at 2 Hz frequency resolution of all observations in 
which the 450 Hz QPO was detected is shown in the bottom trace. The lower 
frequency $\sim 18$ Hz QPO can also be seen.

\vskip 10pt

Figure 3: Power spectra computed from the three
observations in which we detected the 450 Hz QPO and in which Remillard et al.
reported either detections or evidence of the 300 Hz QPO. The upper spectrum 
was computed from the 2 - 12 keV single bit mode data only while the lower 
spectrum was computed using only the 13 - 27 keV event mode data. Both the 
300 and 450 Hz QPO are clearly detected simultaneously.

\vskip 10pt

Figure 4: The radius of the inner most stable
circular orbit (ISCO) as a function of dimensionless black hole angular 
momentum, $j$, for the mass limits for GRO J1655-40 of Shahbaz et al. 
(1999) and Orosz \& Bailyn (1997) (left panel). The curves denoting the 
ISCO radii are labelled with the corresponding mass (5.5, 6.8, 7.2, or 
7.9 $M_{\odot}$). Also shown for each mass are the 
corresponding radii at which the orbital frequency is equal to 
450 Hz. Curves corresponding to a particular mass are plotted with the same 
line style (ie. solid, dashed, etc.). For example, for a 5.5 $M_{\odot}$ 
black hole one requires $j \sim 0.15$ in order for the Kepler frequency at the 
ISCO to be 450 Hz. 
The Radial epicyclic frequency (curved lines) and the nodal precession 
frequency (diagonal lines) vs Keplerian frequency for the Shahbaz et al. 
mass limits for GRO J1655-40 and for several different values of the 
dimensionless angular momentum $j$ (right panel). the solid, dotted and 
dashed curves correspond to $j = 0.2, \; 0.4$, and 0.6, respectively. 
The curves for $j=0.2$ are labelled with their corresponding masses in order
to indicate the sense of the mass dependence. The difference frequency 
between the 300 and 450 Hz QPO and the low frequency ($\sim 18$ Hz) QPO are 
plotted with diamonds. Note that the error bars are smaller than the symbol 
size.

\vskip 10 pt

\vfill\eject

\begin{figure*}[htb] 
\centerline{\epsfig{file=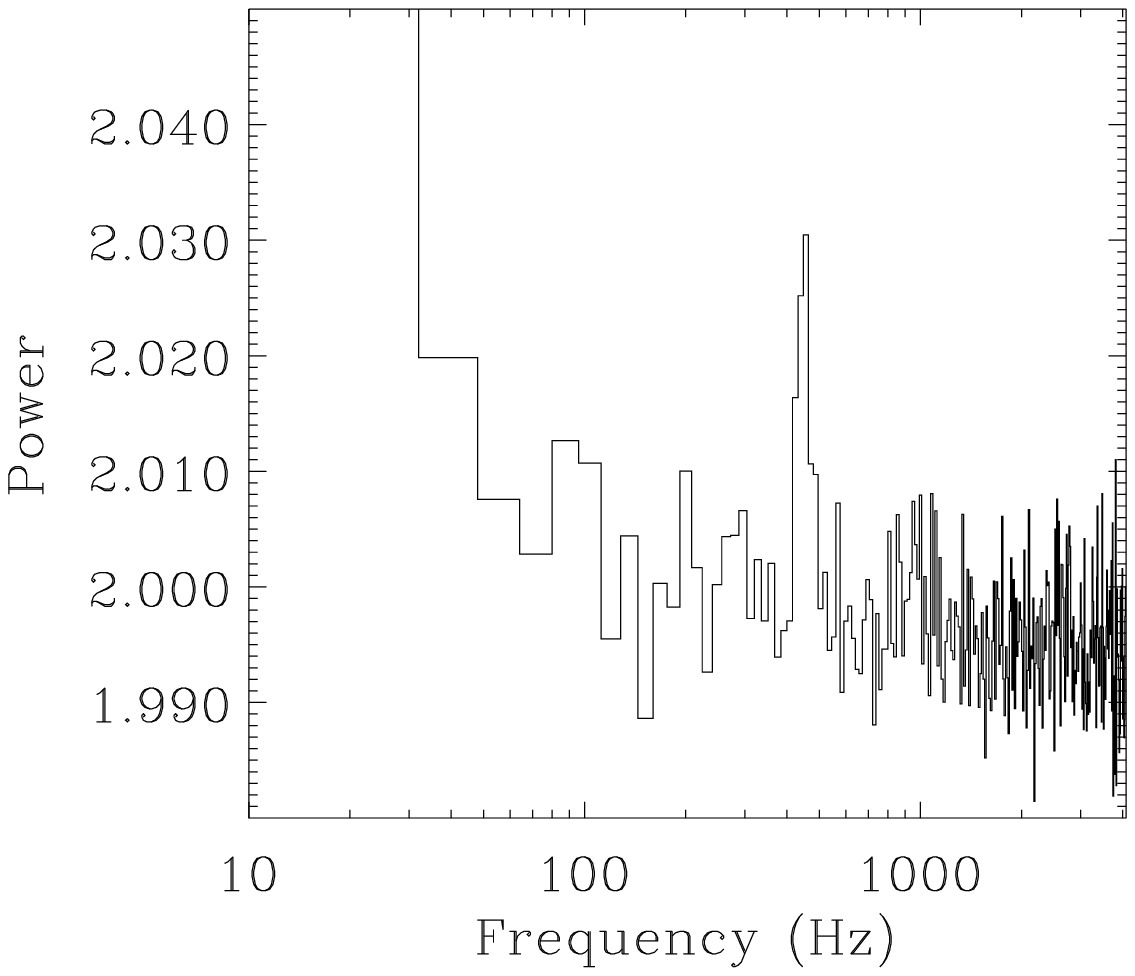,height=6.0in,width=6.0in}}
\vspace{10pt}
\caption{Figure 1}
\end{figure*}

\vfill\eject

\begin{figure*}[htb] 
\centerline{\epsfig{file=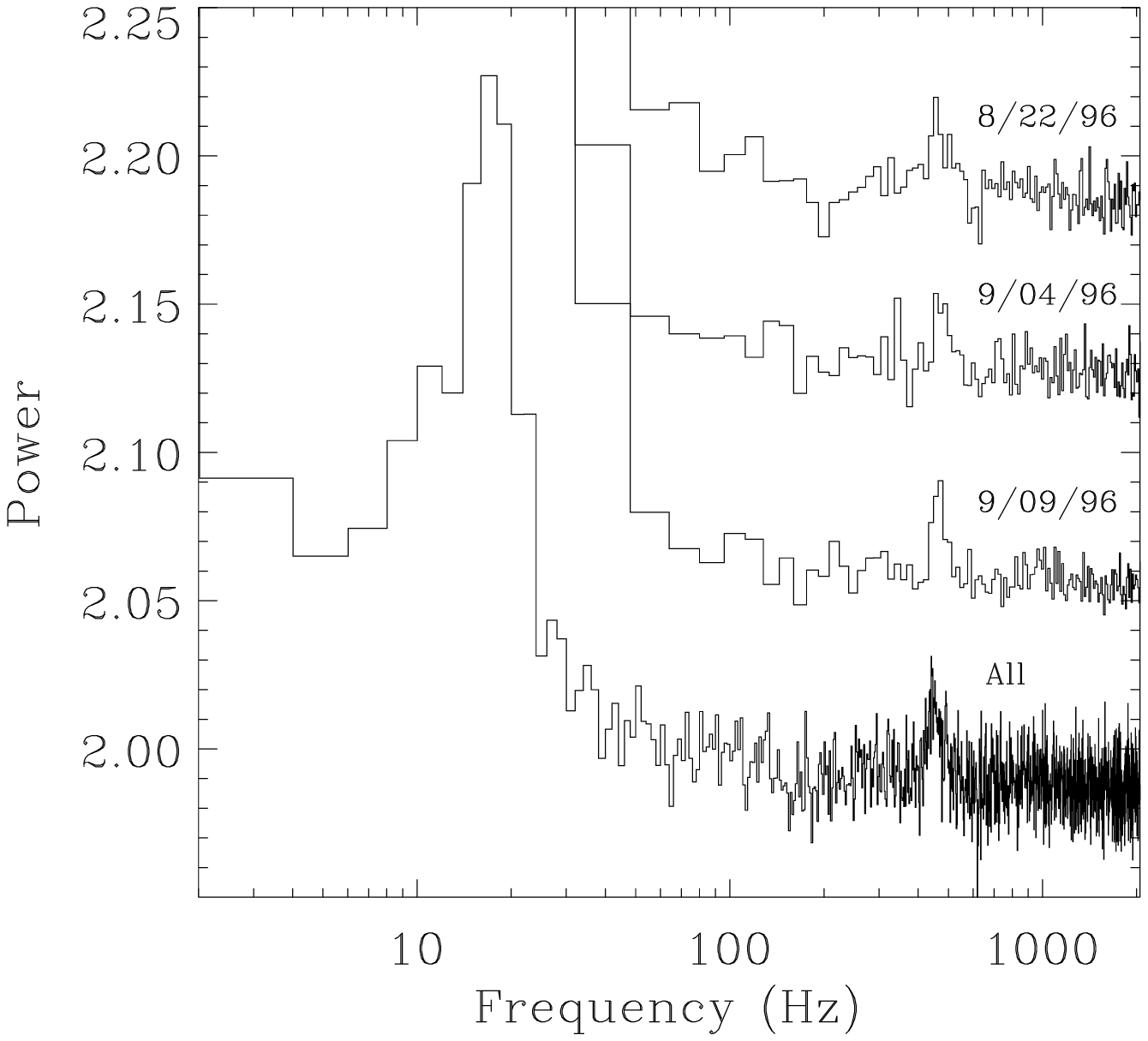,height=6.0in,
width=6.0in}}
\vspace{10pt}
\caption{Figure 2}
\end{figure*}

\vfill\eject

\begin{figure*}[htb] 
\centerline{\epsfig{file=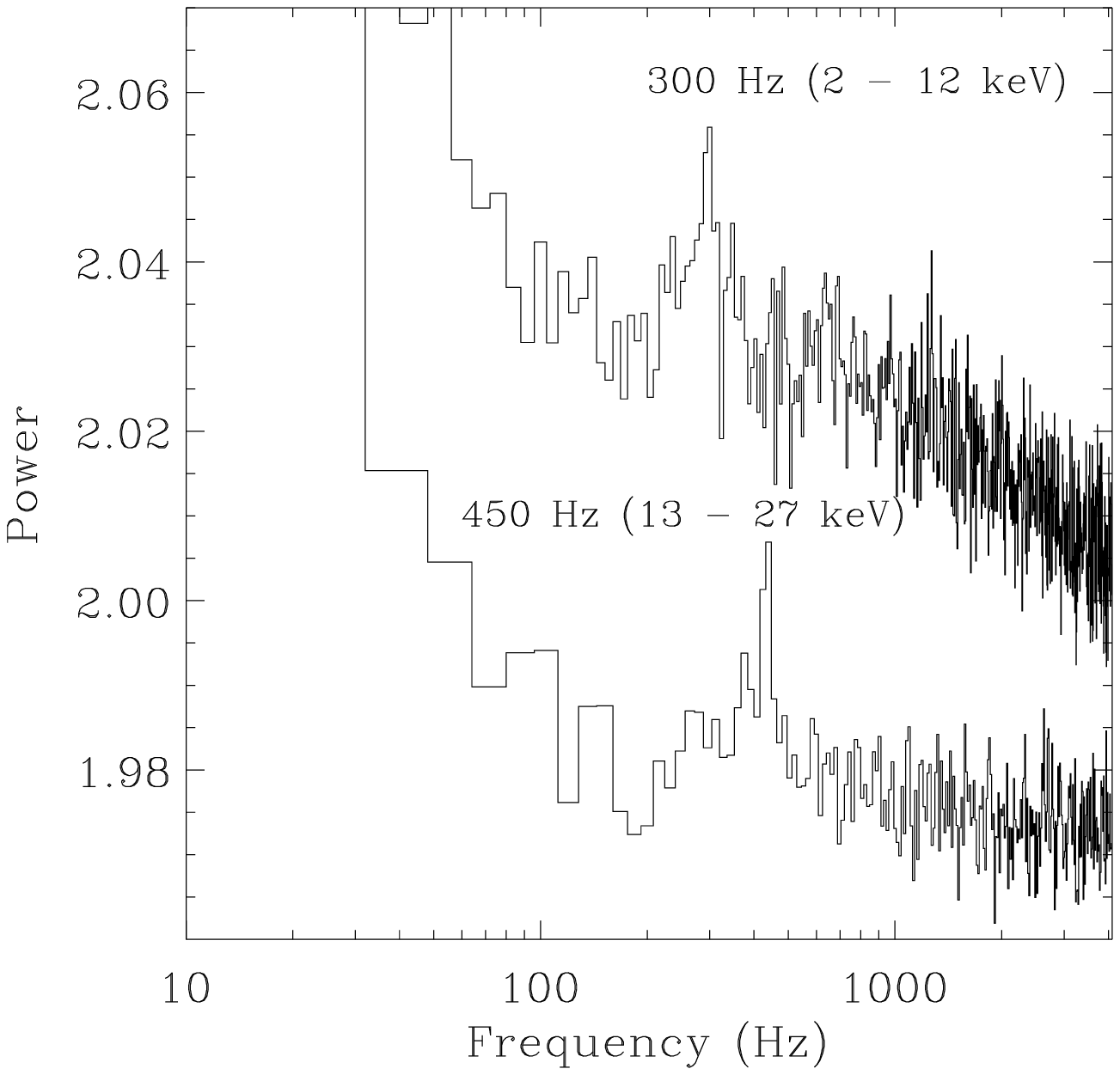,height=6.0in,width=6.0in}}
\vspace{10pt}
\caption{Figure 3}
\end{figure*}

\vfill\eject

\begin{figure*}[htb] 
\centerline{\epsfig{file=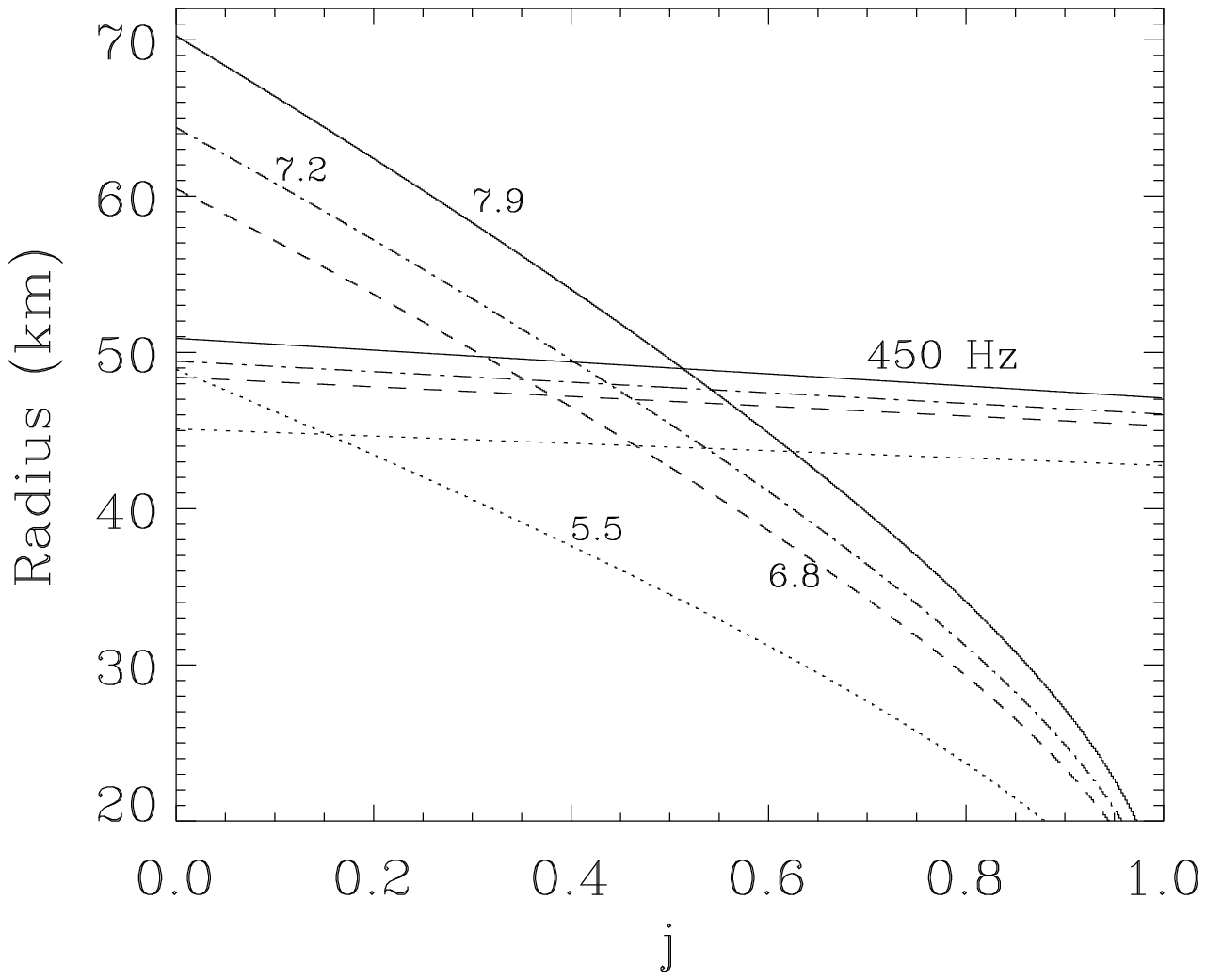,height=6.0in,width=6.0in}}
\vspace{10pt}
\caption{Figure 4a}
\end{figure*}

\vfill\eject

\begin{figure*}[htb] 
\centerline{\epsfig{file=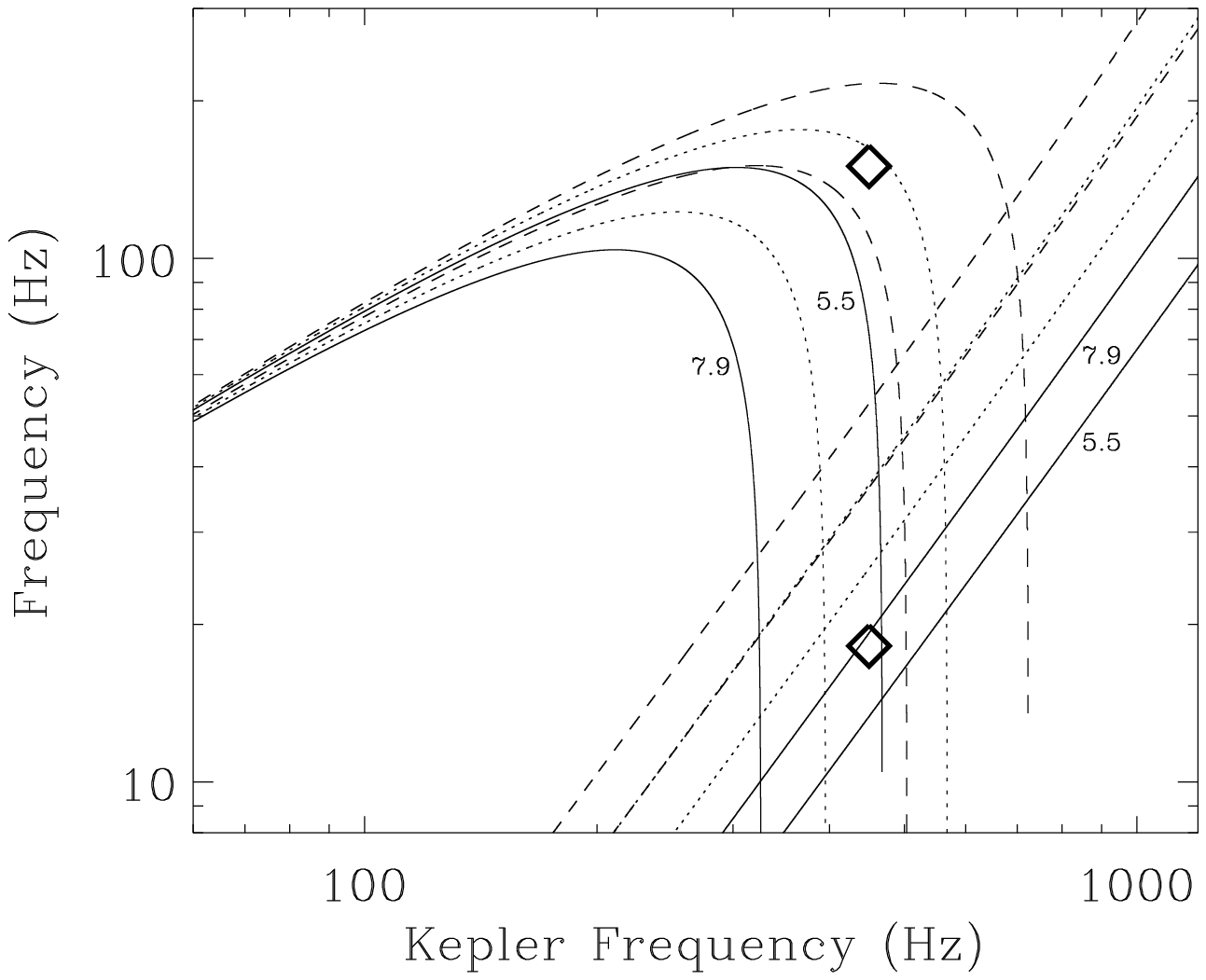,height=6.0in,width=6.0in}}
\vspace{10pt}
\caption{Figure 4b}
\end{figure*}

\vfill\eject

\end{document}